\documentclass[pra,reprint,nofootinbib,showpacs,superscriptaddress,longbibliography]{revtex4-2}
\usepackage{graphicx} 
\usepackage{hyperref}
\usepackage{amsfonts}
\usepackage{amsmath,amssymb}
\usepackage{bm} 
\usepackage{color}
\usepackage{epstopdf} 
\usepackage{epsfig}
\usepackage{subfig} 
\usepackage{float}

{\rm }

\def\be{\begin{equation}}
 \def\ee{\end{equation}}
 \def\bea{\begin{eqnarray}}
 \def\eea{\end{eqnarray}}
 \def\bes{\begin{eqnarray}}
 \def\ees{\end{eqnarray}}
 \def\bi{\begin{itemize}}
 \def\ei{\end{itemize}} 

\def\2{\frac{1}{2}}
\def\4{\frac{1}{4}}

\begin{document}

\title{Radiation-Induced Dark Counts for Silicon Single-Photon Detectors in Space}

\author{Brandon A. Wilson}
\email{wilsonba1@ornl.gov}
\affiliation{Detonation Forensics and Response Group, National Security Sciences Division, Oak Ridge National Laboratory, Oak Ridge, TN 37831, USA}

\author{Alexander Miloshevsky}
\affiliation{Quantum Information Science Group, Computational Sciences and Engineering Division, Oak Ridge National Laboratory, Oak Ridge, TN 37831, USA}
\affiliation{Department of Mechanical, Aerospace, and Biomedical Engineering, University of Tennessee, Knoxville, TN 37996, USA}

\author{David A. Hooper}
\affiliation{Detonation Forensics and Response Group, National Security Sciences Division, Oak Ridge National Laboratory, Oak Ridge, TN 37831, USA}


\author{Nicholas A. Peters}
\affiliation{Quantum Information Science Group, Computational Sciences and Engineering Division, Oak Ridge National Laboratory, Oak Ridge, TN 37831, USA}
\affiliation{Center for Interdisciplinary Research and Graduate Education, The University of Tennessee, Knoxville, Tennessee 37996, USA}

\date{\today}

\begin{abstract} 
Single-photon detectors operating on satellites for use in a quantum communications network can incur large dark count rate increases from the natural radiation environment of space. Displacement damage to the material lattice of a detector from the ionizing radiation can result in a permanent dark count increase in the detector. In this work, we analyze the radiation-induced dark count rate of a silicon single-photon avalanche diode onboard a satellite at different orbiting altitudes, as 
 well as, the additional radiation from a nuclear-disturbed environment caused by a high-altitude nuclear explosion. For detectors on low Earth orbit satellites, protons are the biggest source of radiation damage and are best mitigated by choosing an orbit that minimizes exposure when passing through the South Atlantic Anomaly and Polar Cusps. Detectors on medium Earth orbit and geostationary orbit satellites, if shielded by more than 10 mm of aluminum, provide the best platform in terms of the least amount of radiation damage to the detectors. In the event of a high-altitude nuclear explosion, the artificial radiation belts produced by the explosion will cause too much damage to silicon single-photon detectors on low Earth orbit satellites and render them unfit for quantum communications in less than a day. Higher orbit satellites  will only suffer minor dark count rate increases from the artificial radiation belts.\footnote{This manuscript has been authored in part by UT-Battelle, LLC, under contract DE-AC05-00OR22725 with the US Department of Energy (DOE). The US government retains and the publisher, by accepting the article for publication, acknowledges that the US government retains a nonexclusive, paid-up, irrevocable, worldwide license to publish or reproduce the published form of this manuscript, or allow others to do so, for US government purposes. DOE will provide public access to these results of federally sponsored research in accordance with the DOE Public Access Plan (http://energy.gov/downloads/doe-public-access-plan).}

\end{abstract}

\maketitle

\section{Introduction}
\label{sec:1}

Quantum communications between ground stations and satellites as well as between satellites themselves, offer the possibility for near-term long-distance quantum networking. The quantum network would employ free-space optical links over which quantum bits, or ``qubits'', are encoded on individual photons. These systems suffer from large (typically $\geq$20 dB) optical channel losses \cite{aspelmeyer2003long, bonato2009feasibility,liao2017satellite} due to absorption, scattering, and diffraction of the photons and unlike traditional communications, the optical signal from a quantum link cannot be amplified without adding an unacceptable amount of noise due to the non-cloning principle \cite{wootters1982single}. To detect the faint signals of the optical link, a detector system must have a high detection efficiency and a low background noise. The background noise is a combination of ambient light in the detector's mode collection (i.e., the mode collected by the optical system, including  wavelength, timing, and spatial modes) and intrinsic detector noise, which is typically defined as a dark count rate (DCR). For the silicon detectors considered here, a dark count is a false-positive single photon detection typically caused by a spurious charge, typically originating from thermally generated carriers \cite{kim2011ultra}. Silicon single-photon detectors are thus typically cooled to help reduce the DCR. Another mechanism that contributes to the DCR of a detector is the damage caused to the detector's material lattice by ionizing radiation. Ionizing radiation can cause defect centers in the detector material which trap charge carriers and increase the level of both thermally generated carriers and band-to-band tunneling in the semiconductor material~\cite{xu2017comprehensive}. A general global quantum communications network will need single-photon detectors in satellites which would subject the detectors to the natural background radiation of space. High-fidelity quantum communications are highly desirable, resulting in the need to minimize radiation-induced dark counts. In this work, we investigate the cumulative effects of radiation on satellite-borne silicon single-photon avalanche diodes from both natural and nuclear-disturbed space environments caused by a high altitude nuclear explosion (HANE). We calculate the dark count impact and strategies for reducing it, including shielding and low-temperature operation. In addition, we analyze the impact to the DCR of different satellite orbits and their radiation environments.  

\section{Background}
\label{sec:2}
\subsection{Single Photon Detectors}
\label{qdetector}
Single-photon detectors used for quantum communications need to have high detection efficiencies, low timing jitter and low dark count rates to ensure high fidelity system operation. This is even more crucial for a ground to satellite connection which typically has large optical channel losses  \cite{aspelmeyer2003long,bonato2009feasibility,liao2017satellite}. Additionally, for satellite borne quantum detectors, the detector must deal with the cost ramifications of size, complexity, power usage and cooling onboard the satellite. Single-photon avalanche diodes (SPADs) are very attractive for satellite applications as they do not require the cryogenic overheads of their main competitor, superconducting nanowire single-photon detectors. SPADs are a mature technology and have a precedent of being used in satellites for different applications, including quantum communications \cite{yang2019spaceborne}.

SPADs are semiconductor-based devices that operate with a p-n junction that is sensitive to both ionizing radiation (gamma rays, protons, etc.), along with the vast majority of the visible and near infrared spectrum. The detector is reverse biased above its breakdown voltage so that the liberation of one electron (from a photon or other particle) causes the free electron to accelerate quickly and with enough energy to ``knock'' other bound electrons out, causing an ``avalanche effect.'' The result is a substantial collection of charge in the diode \cite{eisaman2011invited}. To stop the avalanche process, the detector must be quenched by lowering the bias voltage below the materials breakdown voltage. Once the detector is quenched, the bias voltage is brought above the breakdown voltage and the detector is operational again. Additional voltage above the breakdown voltage can improve the detection efficiency of the detector but at the cost of increasing the DCR \cite{yang2019spaceborne, sun1997measurement}. Radiation induced damage intensifies the increased DCR from excess bias voltage so that even with increased detection efficiency, the fidelity of the measurement decreases (the signal increase is not as big as the resulting background increase) \cite{yang2019spaceborne}. For this work, we assume the bias voltage was concurrent with the breakdown voltage. SPADs can be made from silicon, germanium, InGaS and other III-V bandgap materials, but silicon is the most common. For the analysis in this work, we consider the performance of a silicon SPAD configured as a commercial off the shelf detector (for example, Execelitas' super-low K factor silicon APD).

Besides excess bias voltage, the dark count rate in a silicon SPAD is also directly related to its operating temperature. Multiple researchers have documented this phenomenon \cite{sun1997measurement,anisimova2017mitigating,tan2013silicon,sun2004space} and noticed that the dark count rates have an exponential relationship to temperature (see Eq.~1). The DCR at temperature $T$ is related to the DCR at a specific temperature $T_0$ multiplied by the exponential of the temperature difference. The constant $\alpha$ is experimentally determined and is unique to the detector.  
\begin{equation}
DCR(T)=DCR(T_0) e^{-\alpha (T_0 -T)}
\label{eq:DCR}
\end{equation}

It is generally expected that the DCR should change by a factor of 10 for every 20~$C^{\circ}$ change in temperature \cite{nightingale1990new}. Whether Eq. \ref{eq:DCR} holds up below -100~$C^{\circ}$ is unknown as -100~$C^{\circ}$ is the lowest temperature that was experimentally tested \cite{sun1997measurement,anisimova2017mitigating,tan2013silicon,yang2019spaceborne}. For our analysis, we used  $\alpha = -0.09757$, as experimentally determined in reference~\cite{yang2019spaceborne}. The detector used in reference~\cite{yang2019spaceborne} was an Execelitas super-low K factor (SLiK) silicon avalanche photodiode, which is an example of the current silicon SPAD industry standard.

In lowering the operating temperature of a silicon SPAD, the DCR will reduce but the afterpulsing noise will increase.  Afterpulsing is the delayed release of a trapped charge (trapped by defects in the material) that is released shortly after the initial avalanche event (with an exponential time release characteristic). The afterpulsing phenomenon is dependent on material type, temperature, number of defects and charging times \cite{ziarkash2018comparative}. Lowering the operating temperature in a SPAD increases the afterpulse probability \cite{gundacker2020silicon}. For this work, we are analyzing a silicon SPAD and so afterpulsing will likely occur on a timescale of less than 50 ns \cite{hadfield2009single} and the afterpulse tail dies off after roughly 100 ns \cite{ziarkash2018comparative}. Using an afterpulse-blocking time of around 100 ns should make the afterpulsing in a silicon detector system negligible and thus, they are not considered further.  For InGaAs detectors and 1550 nm optical links, this is not the case as they are more susceptible to afterpulsing and the pulses have a longer duration (time scale on the microseconds \cite{ziarkash2018comparative}); future modeling of the radiation effects on InGaAs SPADs will need to address the afterpulsing/DCR temperature dilemma.

\begin{figure}[bht]
	\includegraphics[width=.45\textwidth]{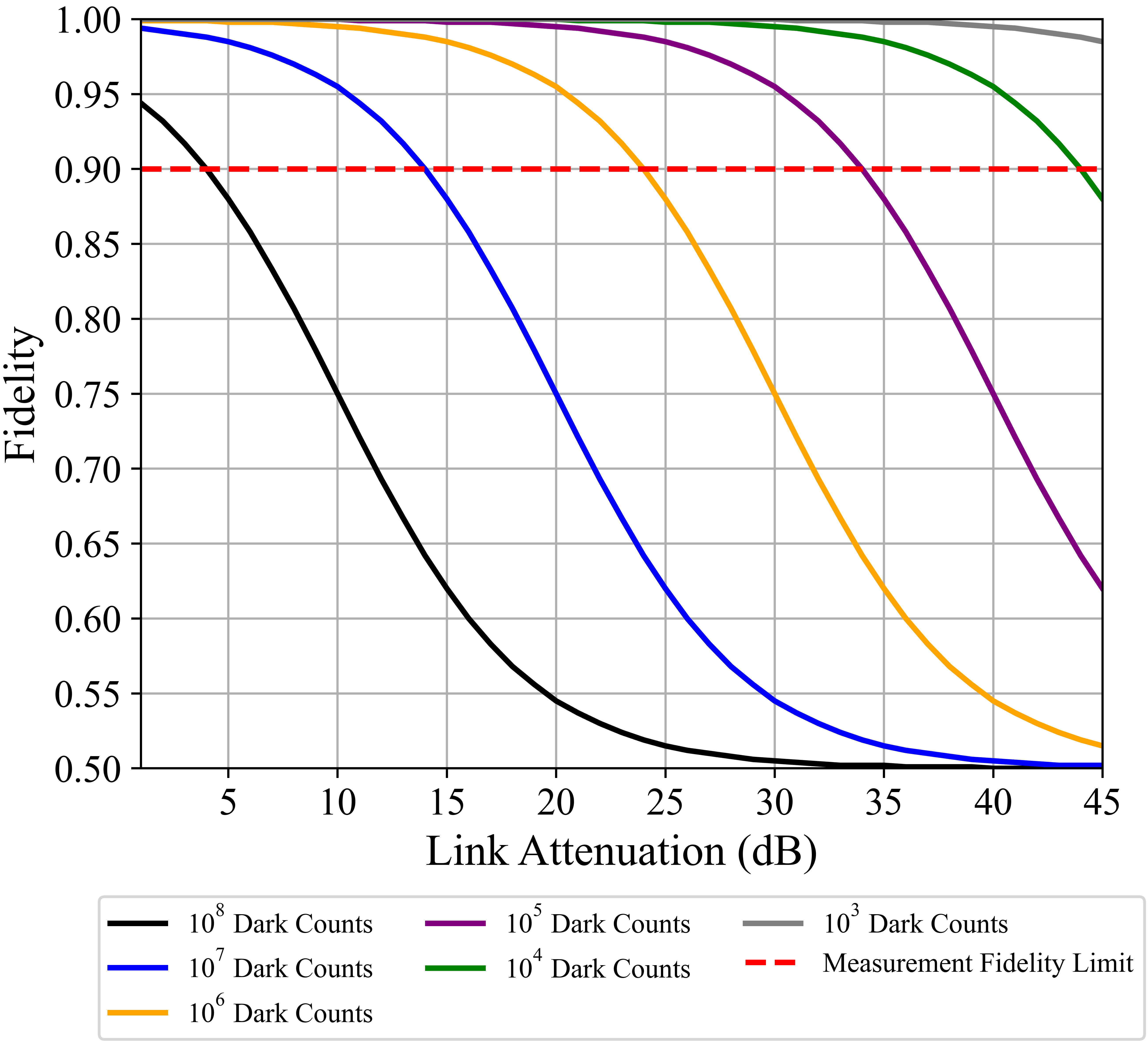}
	\captionsetup{justification=raggedright,
					singlelinecheck=false }
	\caption{Example calculation of the fidelity upper bound for a quantum link  as a function of link attenuation for several different detector dark count rates.}
	\label{fig:1}
\end{figure}

Radiation-induced dark count rate increases will affect the fidelity of the link performance. Generally, in the context of quantum communications, system noise sources such as dark counts, have the effect of transforming pure states ($|\phi_1\rangle$) into  mixed states ($\rho_2$). The fidelity ($F$) is a measure of overlap between two quantum states and is calculated as $F=|\langle\phi_1|\rho_2|\phi_1\rangle|$. The fidelity is commonly used as a figure of merit for quantum communications protocols~\cite{liang2019quantum}. If a transmitted state and a received state are the same, the fidelity is one. If one measures random single qubit noise, the fidelity with any pure state will be $0.5$. As an example, for a quantum channel defined by the transmission of an ideal single photon, operating at a 10 MHz repetition rate and with a detector timing gate of 1~ns, the fidelity as a function of different dark count rates and total system attenuation is shown in Fig. \ref{fig:1}. We assume a fidelity of 0.9 (dashed red line) as a rough lower bound for the minimum fidelity required for quantum communications~\cite{wang2015quantum,BassoBasseteabe6379}. 
 In terms of the quantum bit rate error (QBER), the fidelity and error rate are closely associated for a system with the QBER being the probability of error while the fidelity is the state overlap. In this case, one minus the fidelity is the QBER. For BB84 and SARG04 quantum key distribution protocol, the maximum QBER that can be tolerated is around 11 percent \cite{shor} and 10-15 percent \cite{branciard} respectively, which is equivalent to approximately 0.85-0.9 fidelity. 
 
\subsection{Radiation Effects on Single-Photon Detectors}
\begin{figure}[ht]
	\includegraphics[width=.45\textwidth]{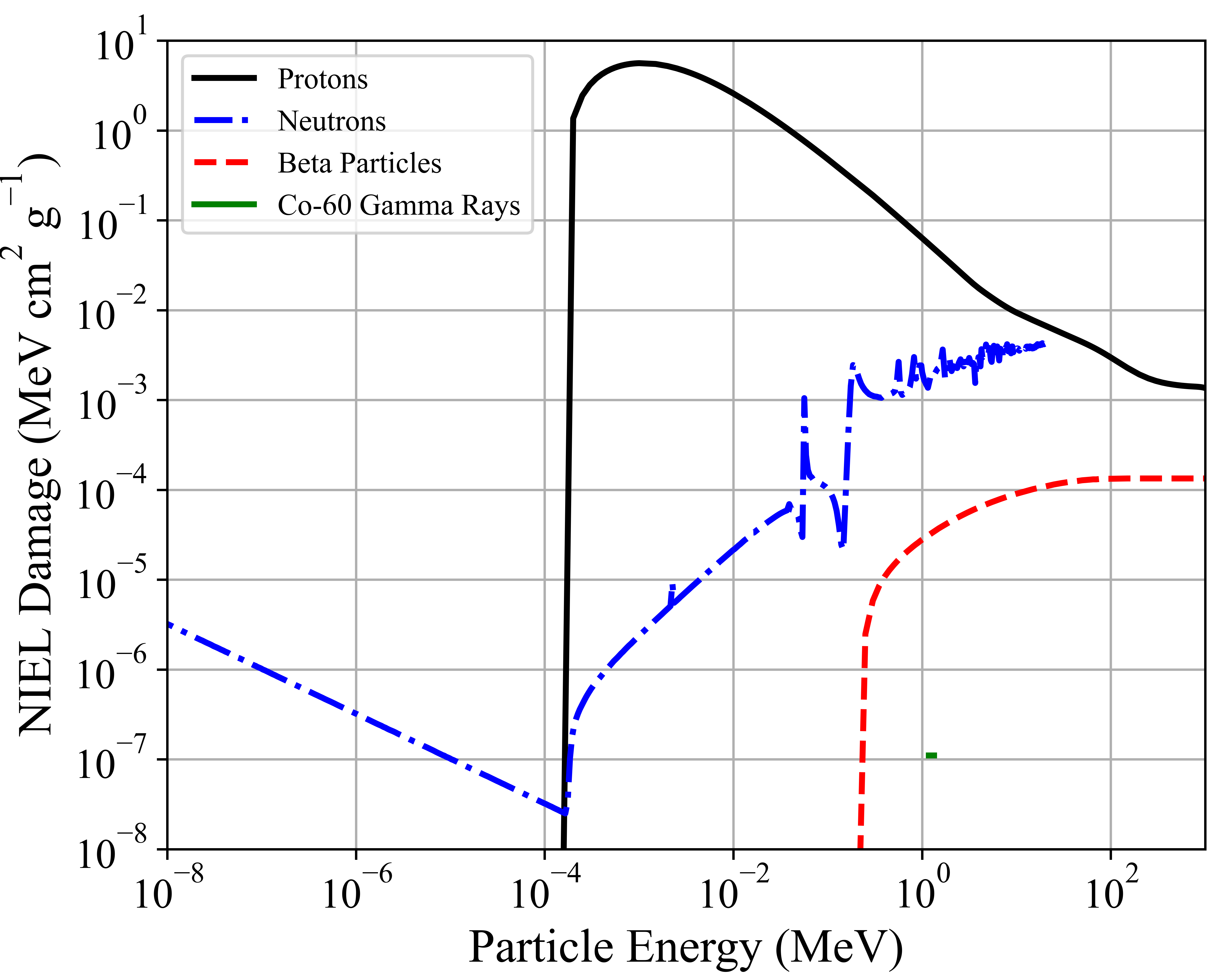}
	\captionsetup{justification=raggedright,
					singlelinecheck=false }
	\caption{NIEL damage in silicon for protons, neutrons, beta particles and $^{60}$Co gamma rays as a function of the particle energy assembled from \cite{jun2003proton,summers1993damage,el2018gamma,ougouag1990differential}.}
	\label{fig:2}
\end{figure}

Ionizing radiation in the form of gamma rays, beta particles, neutrons and protons can cause damage to the lattice of a semiconductor through ionization and displacement damage. Ionization damage is the liberation of an electron from an atom in the material while displacement damage is the dislodgment of an atom from its location in a material's lattice. An ionization event would likely trigger the avalanche response of the detector resulting in a false (dark) count. Displacement damage from radiation particles will not only trigger false counts but also create permanent defect locations in the semiconductor crystal (vacancies, interstitials, etc.) that will create new energy levels in the material's bandgap \cite{srour2000universal}. These new energy levels cause a permanent increase in the dark count rate of the detector. Radiation-induced displacement damage is the biggest factor in the permanent DCR increase in SPADs \cite{srour2000universal}. The amount of displacement damage caused by a particle in a semiconductor material is related directly to the non-ionizing energy loss (NIEL) that the particle imparts on the material. NIEL is a measure of the non-ionizing events, the elastic and inelastic collisions between the incident particle and the nuclei of the atoms in the material. NIEL damage in silicon as function of particle type and energy is shown in Fig. \ref{fig:2} (data assembled from these references: proton \cite{jun2003proton}, beta \cite{summers1993damage}, gamma \cite{el2018gamma}, and neutron \cite{ougouag1990differential}).

Protons impart the most NIEL damage per particle in silicon compared to the other particles of interest in this study. As a result, they are the most impactful in terms of the increased DCR in detectors on satellites. Both beta particles and protons have a minimum threshold energy to inflict NIEL damage in silicon while neutrons, being charge neutral, do not. Gamma ray NIEL has been previously determined by researchers by convolving the Compton secondary electron spectrum, induced by the gamma ray, with the beta particle NIEL functions~\cite{summers1993damage,el2018gamma}. This correlation assumes that a gamma ray only creates displacement damage in a material by Compton scattering an electron with high enough energy to cause NIEL. The correlation of NIEL damage to incident gamma rays has only been experimentally determined with Co-60 gamma rays (1.173 and 1.332 MeV), hence the two data point line in Fig. \ref{fig:2}.  

To determine the DCR increase in a detector from the NIEL damage, a correlation is determined from previous researchers' proton irradiation of SPADs and the resulting DCR increase in their detectors. We use the proton irradiation data from \cite{yang2019spaceborne,anisimova2017mitigating,tan2013silicon} and their dark count rate increases to determine the correlation between NIEL damage and permanent DCR increases. We calculate that roughly 36.26 $(MeV~cm^{2}~g^{-1})$ of NIEL damage equals 1 permanent DCR per second increase in a silicon SPAD. This number is calculated for detectors operating at $-10~C^{\circ}$ and at a bias voltage near the breakdown voltage. As an experimentally determined quantity, the correlation between the NIEL damage and the DCR increase has an inherent uncertainty,  and is worthy of future study.

\subsection{Van Allen Radiation Belts}
\begin{figure}[hbt]
	\includegraphics[width=.45\textwidth]{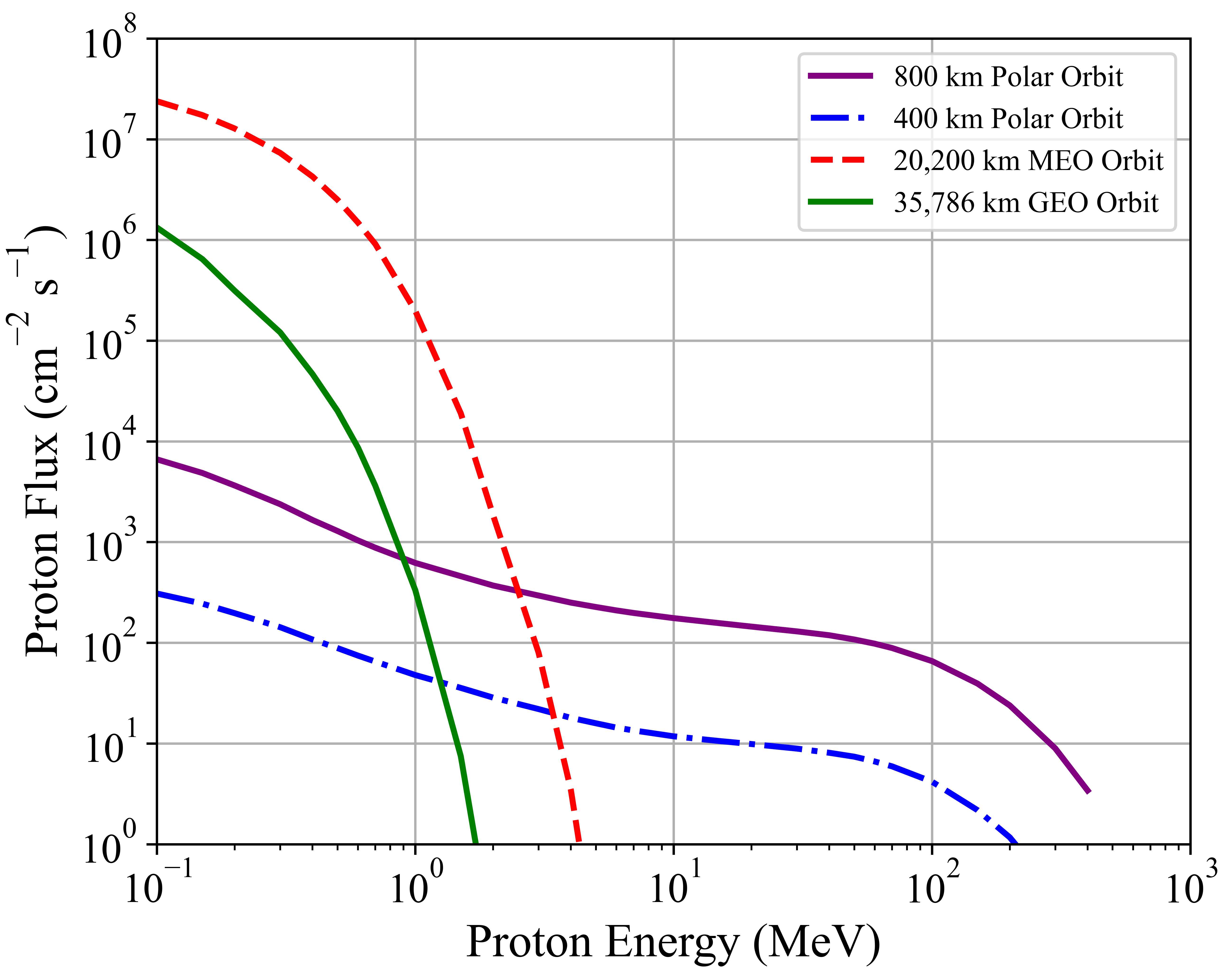}
	\captionsetup{justification=raggedright,
					singlelinecheck=false }
	\caption{Average proton flux (1 year duration) for different satellite orbits \cite{sawyer1976ap,vette1991nasa}.}
	\label{fig:3}
\end{figure}

\begin{figure}[ht]
	\includegraphics[width=.45\textwidth]{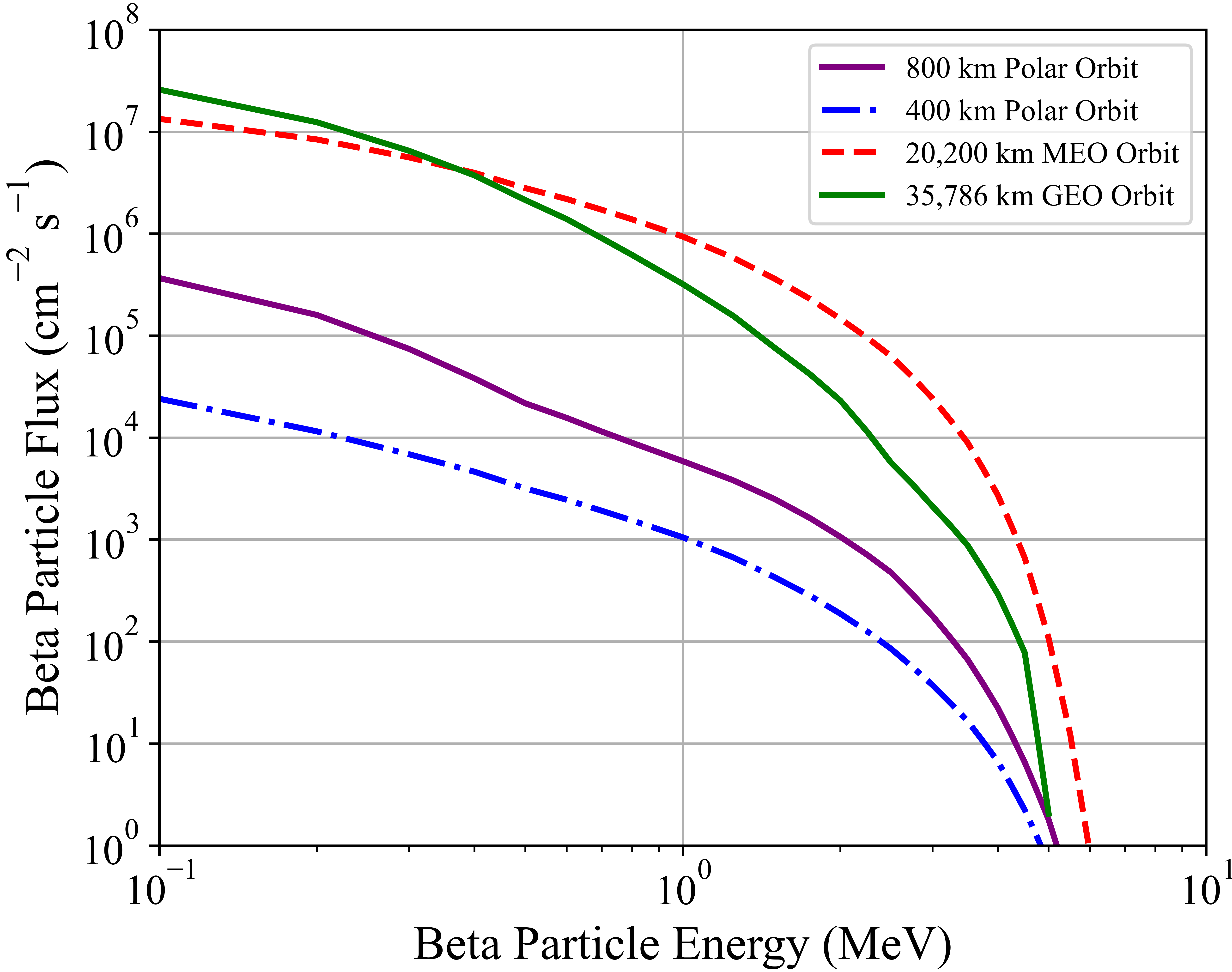}
	\captionsetup{justification=raggedright,
					singlelinecheck=false }
	\caption{Average electron flux (1 year duration) for different satellite orbits.}
	\label{fig:4}
\end{figure}

Quantum detectors on satellites will encounter the largest magnitude of ionizing radiation in the form of protons and electrons as they pass through the different parts of Earth's Van Allen Belts. The Van Allen Belts consist of two toroidal belts: the inner belt which extends approximately from 0.2-2 Earth radii (1,000 -- 12,000 km) and the outer belt which extends from 3--10 Earth radii (19,000 -- 60,000 km). Low earth orbit (LEO) satellites tend to be located below the innermost belt but still travel through parts of the belt as they pass through the South Atlantic Anomaly and the Polar Cusps. Satellites in medium earth orbits (MEO) at 20,200 km and geosynchronous orbits (GEO) at 35,786 km are in the outer belt and will experience a constant bombardment of radiation. The spectrum of the electrons and protons for the different orbits changes significantly between the inner and outer belts. Figs. \ref{fig:3} and \ref{fig:4} depict the average proton and electron flux that each satellite would encounter over the course of a year in space. The data in Figs. \ref{fig:3} and \ref{fig:4} come from the NASA models AP8 and AE8 min \cite{sawyer1976ap,vette1991nasa}. The radiation flux estimates shown in Figs. \ref{fig:3} and \ref{fig:4} are  estimates as the Van Allen Belts change constantly depending on the solar activity. 

The inner belt consists of high energy protons (with up to 400 MeV of energy \cite{sawyer1976ap}) which can penetrate up to 390 mm of aluminum, making shielding difficult. The protons in the outer belt are of low enough energy that they could be effectively stopped by a couple millimeters of aluminum. The electrons in both belts can be shielded effectively, but bremsstrahlung emissions from the beta particles stopped in the shielding are difficult to shield against.   
\subsection{Nuclear Weapon Effects}
In the event of a high-altitude nuclear explosion (HANE), the space radiation environment can change drastically for the various satellite orbits (for the effects of a low-altitude nuclear explosion on a free-space quantum link, see ref\cite{hooper2021effects}). During a HANE event, two major radiation effects can harm a satellite detector: the prompt radiation given off by the exploding weapon (x-rays, gammas and neutrons) and the artificial radiation belts created by the weapon debris beta particles. During a nuclear explosion, prompt x-rays, gamma rays and neutrons stream out of the exploding weapon causing damage to any satellites in the line of sight of the explosion. Due to the near vacuum conditions of Earth's upper atmosphere, the radiation intensity has a $\frac{1}{r^2}$ dependence as air attenuation of the radiation is negligible at these altitudes. The equations for modeling the neutron and gamma fluence are taken from \cite{conrad2010collateral} and are shown in Eqs. \ref{eq:N_flux} and \ref{eq:G_Flux}. The variable $Y$ is the yield of the detonation in Megatons (MT) and $R$ is the distance from the explosion in kilometers (km).
\begin{equation}
    \label{eq:N_flux}
    \phi_{n}=1.6 \times 10^{15} \frac{Y}{R^2}~(\frac{n}{cm^2})
\end{equation}
\begin{equation}
    \label{eq:G_Flux}
    \phi_{\gamma} =9.9 \times 10^{14} \frac{Y}{R^2}~(\frac{\gamma}{cm^2})  
\end{equation}

For a detonation at 100 km, LEO satellites are at the highest risk of damage since they are much closer than MEO and GEO satellites; however, LEO satellites may have the advantage of a reduced field of view to the weapon blast due to Earth shielding the line of sight from their low orbit altitude. The prompt gammas and neutrons traverse through most satellite shielding material and the only way to reduce potential harm is being a greater distance away from the blast. For this study, we are only interested in the radiation effects on a satellite detector and so we ignore the other prompt effects from a HANE that can damage a satellite such as photon induced thermomechanical effects, system generated electromagnetic pulse, x-ray damage to the optics, instrumentation, and solar panels, and weapon debris implantation into the satellite. 

\begin{table*}
    \centering
    \caption{HANE artificial belts 1 hour post detonation for a 1 MT explosion at 100 km.}
    \label{tab:rad_belt}
    
    \begin{tabular}[c]{ |c||c|c|c|} 
        \hline
        Satellite & LEO Polar Orbit & MEO Orbit & GEO Orbit \\ 
        \hline
        \hline
        Latitude (North America) & 5 & 51.2 & 57.9 \\ 
        L-Shell & 1.125 & 4.167 & 6.58 \\ 
        Volume ($m^3$) & $1.17\times 10^{17}$ & $2.54\times 10^{20}$ & $1.77\times 10^{21}$ \\ 
        Beta Particle Flux ($cm^{-2} s^{-1}$) & $4.66\times 10^{8}$ & $1.11\times 10^{5}$ & $1.26\times 10^{4}$ \\
        \hline
    \end{tabular}
\end{table*}

The second HANE environment radiation mechanism that causes damage to a single-photon detector on a satellite is the creation of artificial Van Allen Belts from the weapon debris. The fission products produced from a nuclear explosion emit high energy beta particles that become trapped in Earth's magnetic field. The beta particles are constrained by the Lorentz force and move along Earth's magnetic field lines, occupying the magnetic flux tube that the explosion and debris inhabit \cite{conrad2010collateral}. 
The beta particles oscillate around the magnetic field lines and slowly diffuse around the world to form an artificial radiation belt in the shape of a toroid. The complication of the electromagnetic conditions of Earth's magnetic field following a HANE event and the complexity of the weapon debris transport makes beta particle trapping efficiency calculations hard to accurately predict. Additionally, not many HANE nuclear tests were conducted to get a good sample size and develop an exact understanding of the physics. The predicted trapping efficiency of the beta particles ranges between $1\times 10^{-7}$ and 0.10 (i.e., $1\times 10^{-5}$\% and 10\%) depending on the altitude and latitude of the explosion~\cite{cladis1971trapped}. The Starfish Prime test had a trapping efficiency of around 10\% of the total electrons emitted from the weapon debris~\cite{cladis1971trapped}. For this work, we assume the worst-case scenario for the beta particle trapping in the belts is 10\% and that the center of the belt intersects the orbit of the satellite.

In addition to the trapping efficiency, the beta particle flux in this artificial radiation belt depends on the volume of the magnetic flux tube. The volume of the magnetic tube is dependent on the magnetic latitude or ``L-shell'' that the explosion took place in. Table \ref{tab:rad_belt} above shows the volumes of the corresponding radiation belts necessary to impact a specific satellite orbit. The large volumes of artificial radiation belts needed to impact a MEO or GEO satellite reduces the beta particle flux by 3 to 4 magnitudes from that of a belt at LEO orbits. 

The lifetime of the electrons in these artificial radiation belts is dependent on the L-shell that they inhabit. For artificial radiation belts that would affect a LEO satellite, the lifetime is roughly 100 days \cite{cladis1971trapped}. For belts that would affect MEO and GEO satellites, the mean lifetimes are on average 10 and 5 days, respectively \cite{cladis1971trapped}. This indicates that not only is there a higher beta particle flux in the radiation belts that would affect a LEO satellite, but that they also persist for longer periods of time.  

\begin{figure}[ht]
	\includegraphics[width=.45\textwidth]{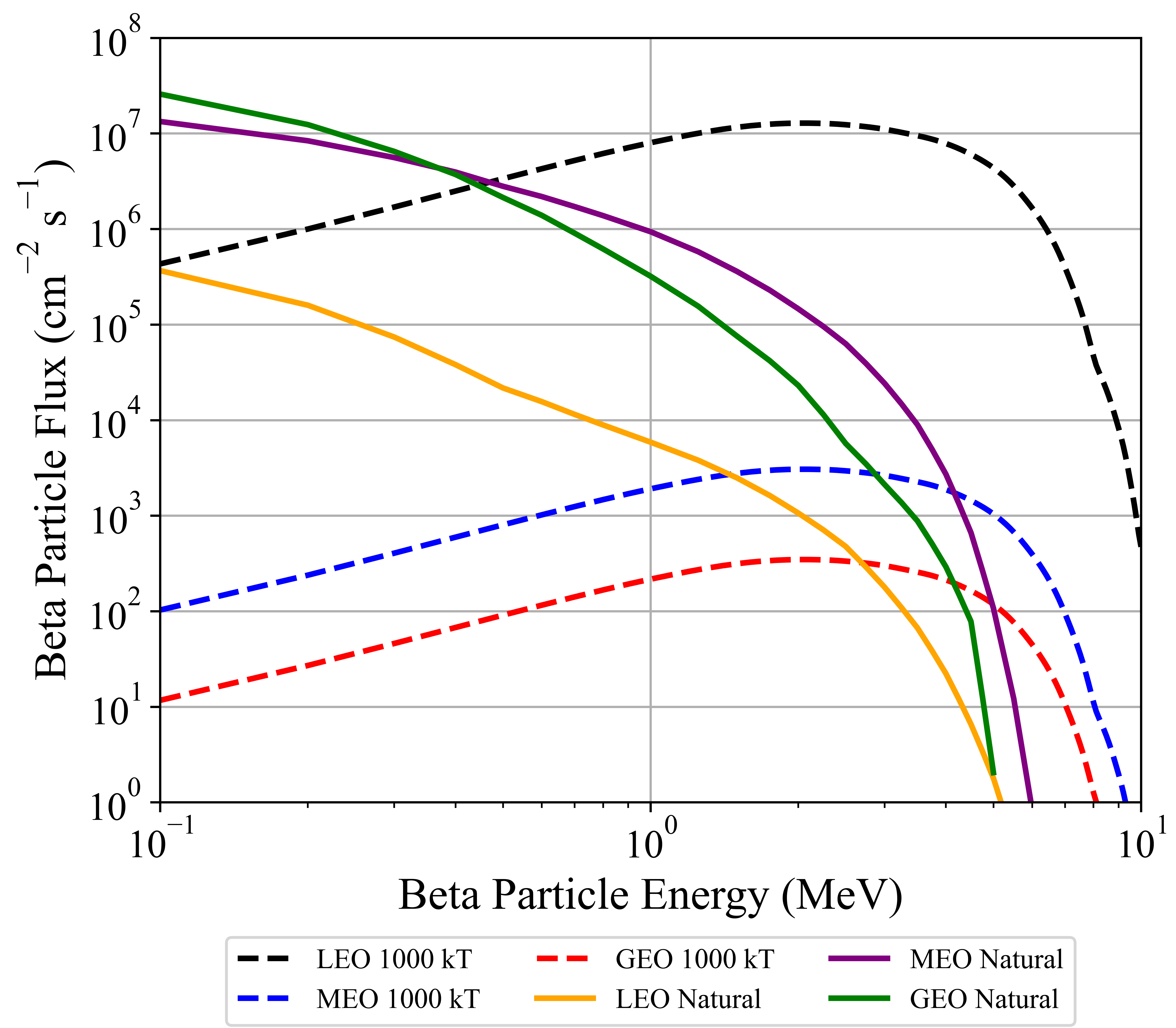}
	\captionsetup{justification=raggedright,
					singlelinecheck=false }
	\caption{Artificial radiation belt beta particle spectrum 1-hour post detonation for different satellite orbits.}
	\label{fig:5}
\end{figure}
The spectrum of beta particles in the artificial radiation belts that are created from a HANE event have a harder spectrum ('harder' spectra have higher particle energies, and 'softer' have lower particle energies) than that of the natural radiation belts. Fig. \ref{fig:5} shows a comparison of the beta particle spectrums and flux can be seen between the different satellite orbits. The data for the beta particle spectrum was taken from the nuclear code package SCALE \cite{scale2020} and represents the beta particle emissions from the fission products of \textsuperscript{235}U over the period of the first 6 hours. The harder spectrum electrons pose a problem in regards to shielding of the satellite detectors. The higher energy beta particles can penetrate up to 20 mm of aluminum shielding, in addition to the creation of high energy bremsstrahlung radiation. The lower energy beta particles in the artificial radiation belt will decay away preferentially compared to the higher energy beta particles, this leads to an even ‘harder’ beta particle spectra in the belts over time \cite{cladis1971trapped}. For this work, we assumed the spectra was constant over time which leads to a slight underestimate of the DCR increase we predict.

\section{results}
\label{sec:4}
\subsection{Shielding}
\begin{figure}[ht]
	\includegraphics[width=.45\textwidth]{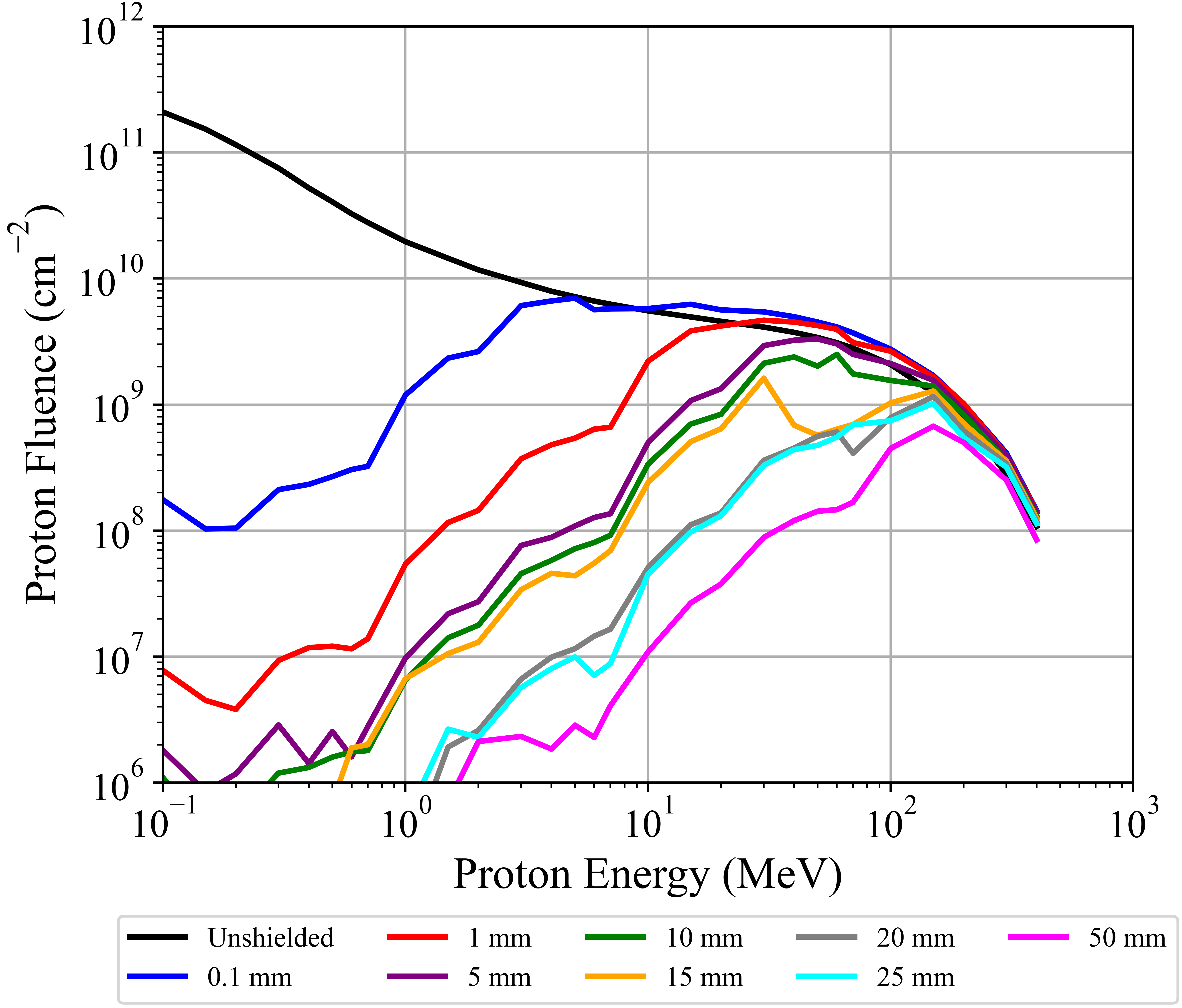}
	\captionsetup{justification=raggedright,
					singlelinecheck=false }
	\caption{MCNP results for the proton fluence through varying aluminum shielding thicknesses for a 800 km polar LEO.}
	\label{fig:6}
\end{figure}

Shielding and cooling of a silicon SPAD detector are the two biggest factors in reducing the radiation damage effects onboard a satellite. Temperature and shielding mechanisms both come with drawbacks, though: satellites are limited on power, space, and weight, so ultra-cold cryogenic coolers or thick radiation shielding on the satellite are likely cost prohibitive. Due to the omnidirectional nature of the radiation flux in the Van Allen belts, radiation shielding must cover the entire detector. For this work, we will assume that the detector is covered with a spherical shell of aluminum and that the detector is connected to the receiver by optical fibers that is routed through the shell to avoid any radiation streaming ports. To model the amount of radiation that impacts a detector as a function of shielding thickness, we use the radiation transport code MCNP \cite{MCNP2018} to model the protons, electrons and the subsequent bremsstrahlung flux through an aluminum shield. The MCNP model uses the proton and beta particle spectra of the Van Allen Belts as the source particles. The proton flux spectrum through various thicknesses of an aluminum shield for an 800 km polar orbiting satellite can be seen in Fig. \ref{fig:6}.
The proton energy spectrum gets harder as the shielding thickness increases and even at 50 mm, protons still penetrate through the shield. The high energy (up to 400~MeV) protons from the lower Van Allen Belt are nearly impossible to shield against without the weight of the satellite becoming cost prohibitive. The protons in the outer Van Allen Belt, which affect the MEO and GEO satellites, have a lower energy range (maximum of 5~MeV) and thus can be completely stopped by only 1 mm of aluminum shielding. Complete proton shielding is possible for MEO and GEO satellites but is likely cost prohibitive for LEO orbits.

\begin{figure}[ht]
	\includegraphics[width=.45\textwidth]{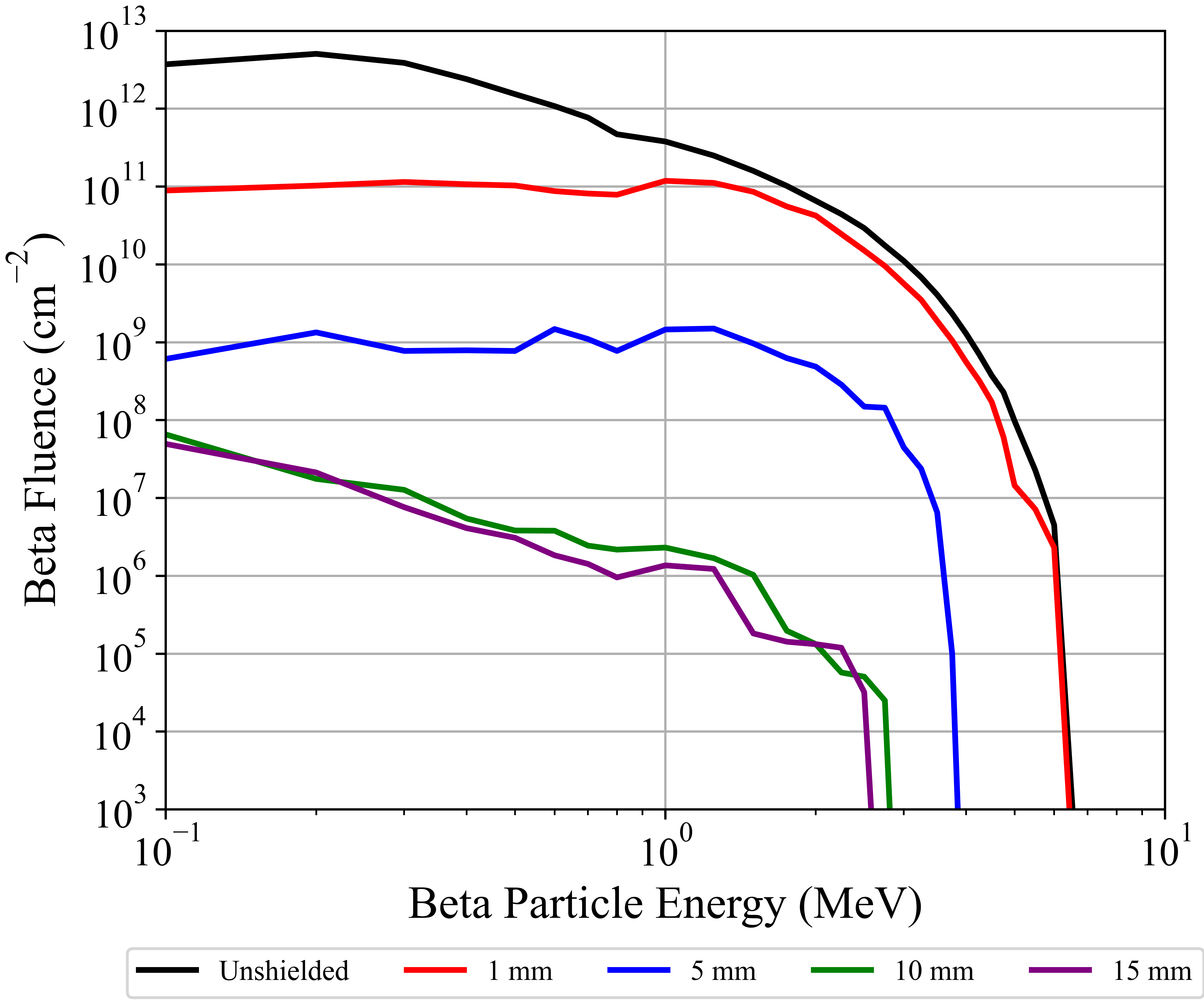}
	\captionsetup{justification=raggedright,
					singlelinecheck=false }
	\caption{MCNP results for the electron fluence through varying aluminum shielding thicknesses for a 800 km polar LEO.}
	\label{fig:7}
\end{figure}

\begin{figure}[ht]
	\includegraphics[width=.45\textwidth]{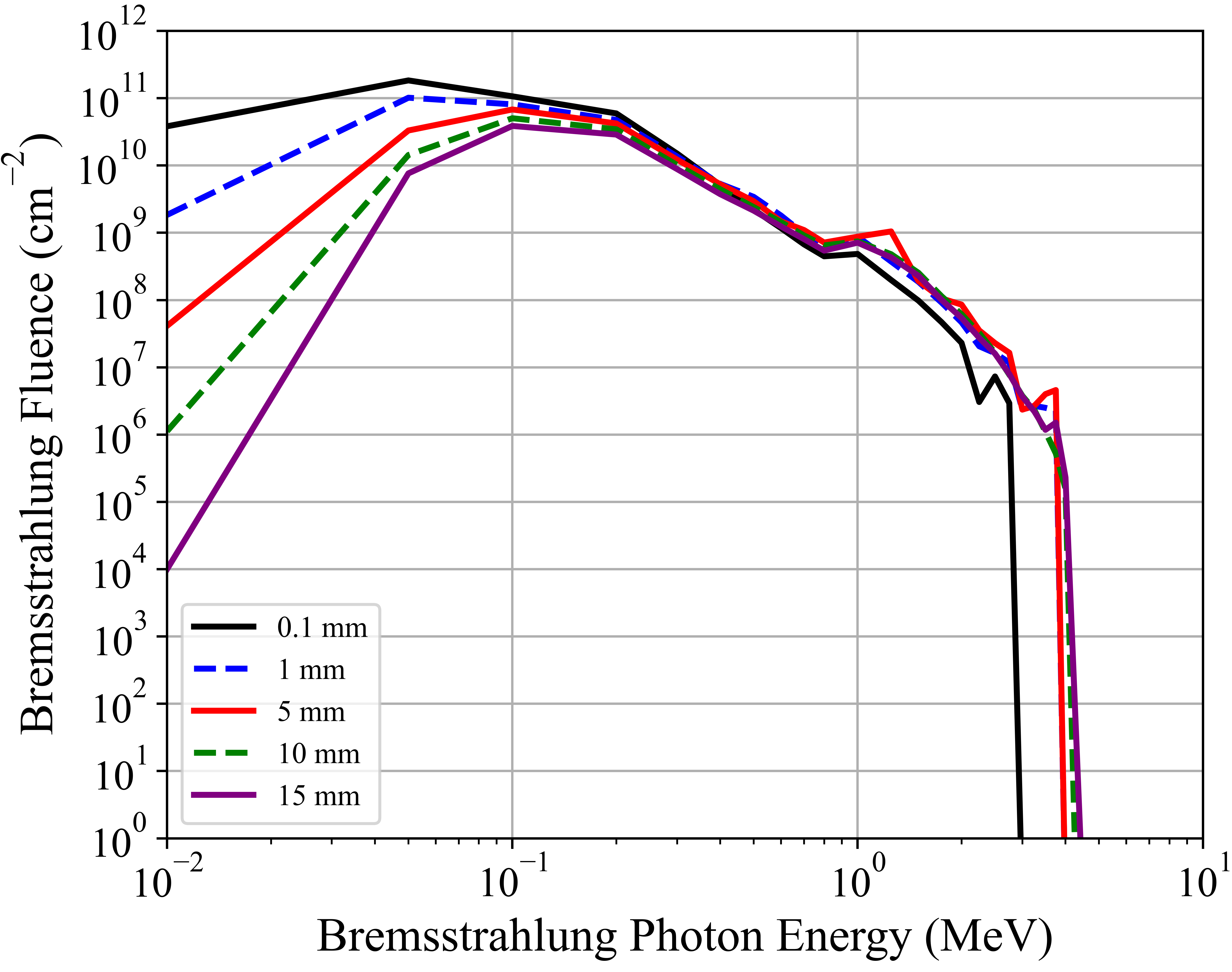}
	\captionsetup{justification=raggedright,
					singlelinecheck=false }
	\caption{MCNP results for the Bremmstrahlung creation and fluence through varying aluminum shielding thicknesses for a 800 km polar LEO.}
	\label{fig:8}
\end{figure}

When shielding energetic beta particles, the ``slowing down'' or deceleration of an electron can result in the creation of electromagnetic radiation (typically in the x-ray to gamma-ray range) that is called bremsstrahlung radiation. The bremsstrahlung spectrum ranges from the incident particle energy down into the low-energy x-ray regime. The higher-energy beta particles thus create higher-energy bremsstrahlung radiation, which includes the gamma-ray energy range. While complete shielding of beta particles is possible for a satellite detector, the resulting bremsstrahlung radiation can still pass through the shielding and impact the detector. Results of the MCNP models for the beta particle shielding and subsequent bremsstrahlung radiation can be seen in Figs. \ref{fig:7} and \ref{fig:8}.

Increasing the shielding around the detectors reduces the beta particle flux and softens the spectrum. Bremsstrahlung photons, however, were only minimally attenuated by increased shielding (except for low energy photons), because x-rays and gammas are less sensitive to low-Z materials like aluminum and require materials with a high density to adequately attenuate them. While Bremsstrahlung radiation is hard to shield against, the displacement damage caused by its gamma rays are orders of magnitude lower than a proton or beta particle. Therefore, a modestly-shielded satellite in MEO and GEO orbit will have its detector incur damage primarily from the Bremmstrahlung gamma rays emitted from the beta particles stopping in the shielding. 

To model the beta particle spectrum that a satellite would encounter in an artificial radiation belt, we assumed a nuclear detonation of 1 MT at an altitude of 100 km and at the magnetic latitude that corresponds to the different satellite orbits. Assuming a worse case scenario of a beta particle trapping efficiency of 10 percent, we model the beta particle spectrum that the satellites would encounter 1-hour post-detonation. The beta spectrum of the artificial belt is much harder than that of the natural belts and requires an aluminum shield of more than 20 mm to fully protect against the beta particles. Beta particle losses in the artificial belts over time tend to occur to the lower energy beta particles at a higher rate than that of the high energy particles \cite{cladis1971trapped}. While this results in an even harder beta particle spectrum (of lower overall intensity) over time, for this work we assume the spectrum remains constant over time. The harder beta particle spectrum from the artificial radiation belt penetrates the shielding more than the natural radiation and also results in higher Bremsstrahlung photons. Similar to the natural belts, when the shielding is increased, the beta particle flux at the detector decreases, but the higher energy Bremsstrahlung photon flux remains relatively constant.

\subsection{Dark Count Rate Increase in Natural Space Environment}
\begin{figure}[ht]
	\includegraphics[width=.45\textwidth]{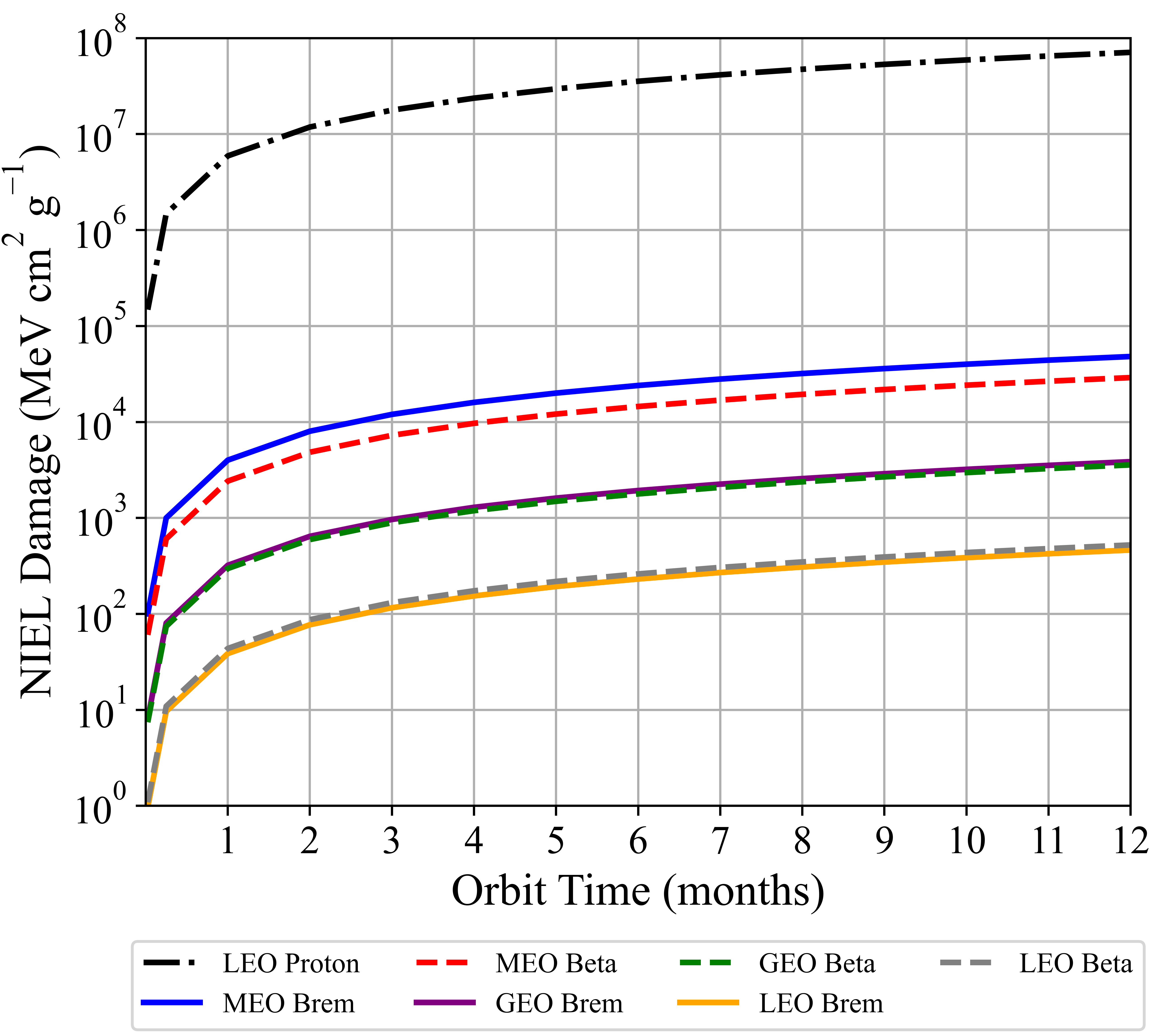}
	\captionsetup{justification=raggedright,
					singlelinecheck=false }
	\caption{NIEL damage to a silicon SPAD versus orbit time for different satellite orbits assuming 10 mm of aluminum shielding. (Brem = Bremsstrahlung)}
	\label{fig:9}
\end{figure}

The radiation flux values obtained through 10 mm of aluminum shielding (calculated in Sec III.A) were multiplied with their respective particle NIEL values (see Fig. \ref{fig:2}) to get a total displacement damage value for the different satellite orbits. The NIEL displacement damage values for the silicon SPAD are plotted in Fig. \ref{fig:9}, with varying orbits and as a function of orbit time.
As expected, the protons cause the largest amount of NIEL damage to the LEO satellite detectors since they are not as easily shielded as the beta particles. When enough shielding is added to reduce the proton fluence by an order of magnitude, the beta particle fluence is reduced by 4--5 orders of magnitude. For the LEO satellite, where proton damage is the dominant mechanism, the more shielding one can put around the detector, the less total damage occurs. For the MEO and GEO satellites, the protons can be completely shielded with only a few millimeters of aluminum and is thus not shown in Fig. 9. At around 10~mm of aluminum shielding, the dominant mechanism for the NIEL damage shifts from the beta particles to the Bremsstrahlung photons. Since Bremsstrahlung photons are difficult to shield against, adding more shielding beyond 10~mm of aluminum only produces minimal NIEL damage reduction and thus is not worthwhile.  

\begin{figure}[ht]
	\includegraphics[width=.45\textwidth]{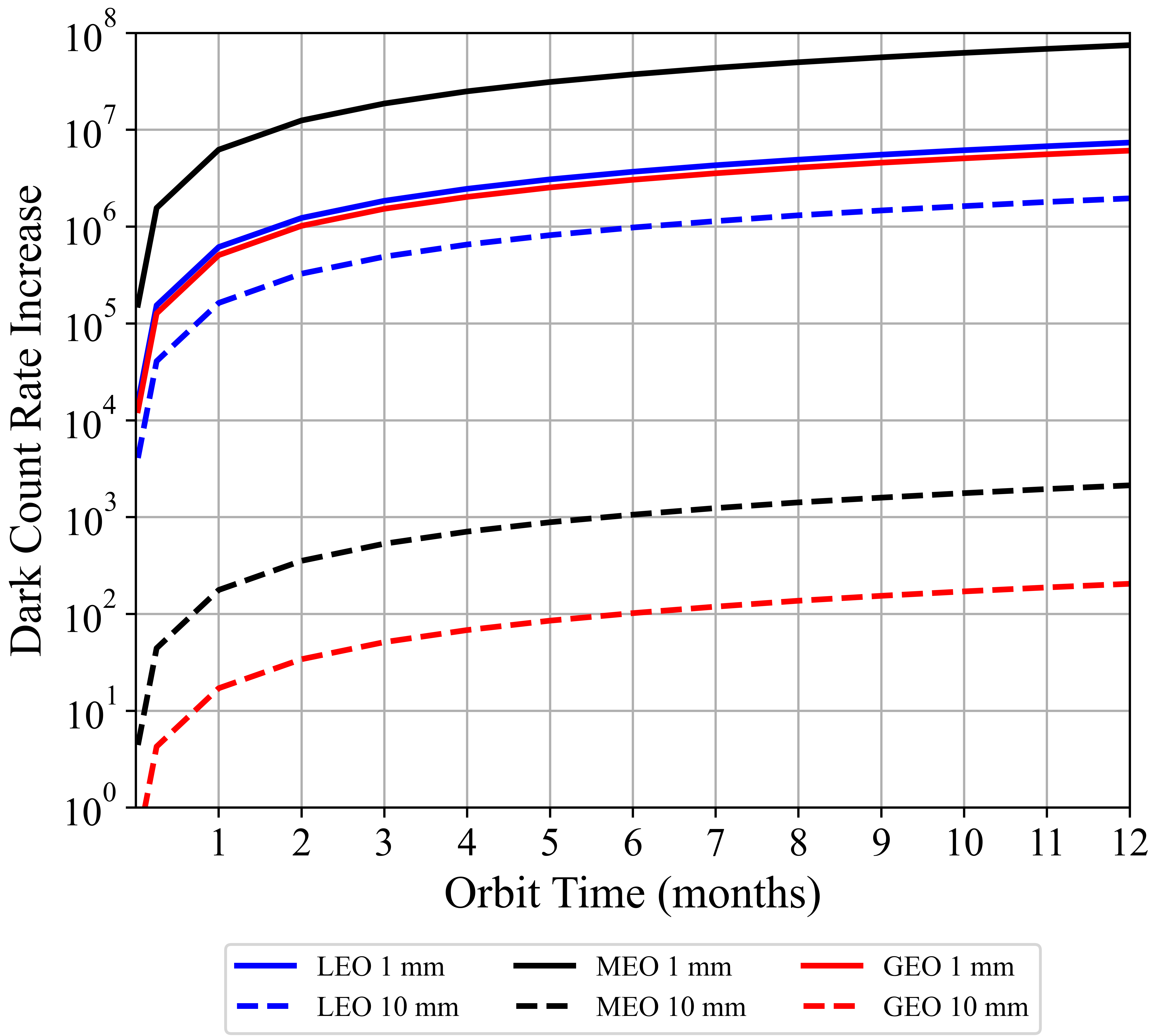}
	\captionsetup{justification=raggedright,
					singlelinecheck=false }
	\caption{Increase in the DCR of a silicon SPAD on a satellite from the natural Van Allen belt radiation.}
	\label{fig:10}
\end{figure}

\begin{figure}[hbt]
	\includegraphics[width=.45\textwidth]{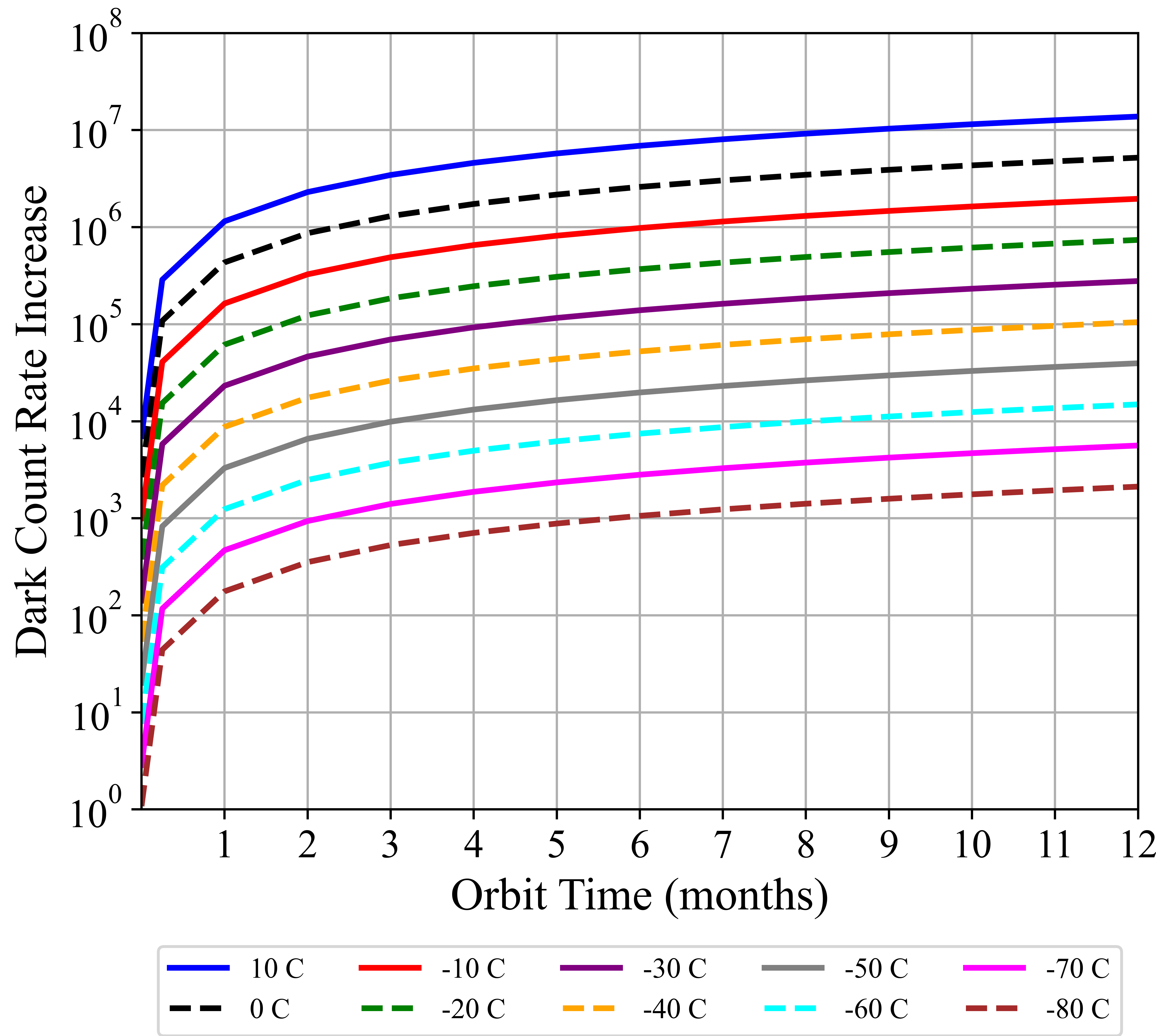}
	\captionsetup{justification=raggedright,
					singlelinecheck=false }
	\caption{Temperature dependence of the radiation induced DCR on a polar LEO at 800 km.}
	\label{fig:11}
\end{figure}

To determine the DCR penalty associated with NIEL damage, a conversion factor of 36.26 $(MeV~cm^{2}~g^{-1})$ per 1 dark count per second increase at an operating temperature of -10~$C^{\circ}$ is applied. The conversion factor was obtained from published experimental literature, though the researchers only considered proton irradiation of silicon SPADs \cite{yang2019spaceborne, anisimova2017mitigating,tan2013silicon}. 
Using this conversion factor, the DCR increase for a silicon SPAD operating at -10~$C^{\circ}$ for the different satellite orbits and shielding is calculated and can be seen in Fig. \ref{fig:10}. The MEO satellite detector incurs the highest DCR increase when minimal shielding is applied, but the DCR drops by 5~orders of magnitude when 10~mm of aluminum shielding is applied due to the almost complete shielding of the beta particles. When thick shielding is applied to a satellite detector ($>10$mm), the LEO satellite incurs the highest DCR increase due to the inability to shield high energy protons in the lower belt.
When the detector temperature is changed, the DCRs change exponentially. By applying Eq. \ref{eq:DCR}, the DCR change as a function of operating temperature for a satellite detector can be determined. Fig. \ref{fig:11} depicts the DCR increase as a function of operating temperature for a 10 mm aluminum-shielded polar satellite orbiting at 800 km.

For quantum communications, it is difficult to carry out many protocols if the fidelity of the transmitted quantum states is not greater than approximately 0.9 (see Fig. \ref{fig:1} and Section \ref{qdetector}). 
To illustrate the situation, we consider an example scenario. Assuming 30 dB of loss in the optical link and a perfect single photon source with a repetition rate of 10 MHz, only 10,000 single photons reach the detector each second.  To achieve 90\% fidelity, the DCR summed over all detector gates would need to be 1,100 counts or less per second (i.e., the total likelihood of an error is $1,100/(10,000+1,100) \approx 10\%$). Assuming a detector  gate time of 1 ns operating at 10 MHz, it would be ``on'' one percent of the time. Thus if it were a free running detector (no timing gates), one would need less than 110,000  dark counts per second. From Fig. \ref{fig:11}, the detector would have to be cooled to -40~$C^{\circ}$ and below in order to achieve a DCR below a hundred thousand.

\subsection{Nuclear Disturbed Environment}
In a nuclear disturbed environment, two effects can cause additional radiation induced DCRs in SPADs: prompt radiation emitted by a weapon and artificial radiation belts created from the weapon debris of a HANE event. The prompt radiation from a nuclear weapon that is of interest in this work consists of gamma rays and neutrons. To estimate the amount of displacement damage that a detector might experience, we use the NIEL damage values from Fig. \ref{fig:2} for neutrons and gamma rays, and Eqs. \ref{eq:N_flux} and \ref{eq:G_Flux} for the distance and yield dependence of the prompt gamma ray and neutron fluxes.
\begin{figure}[ht]
	\includegraphics[width=.45\textwidth]{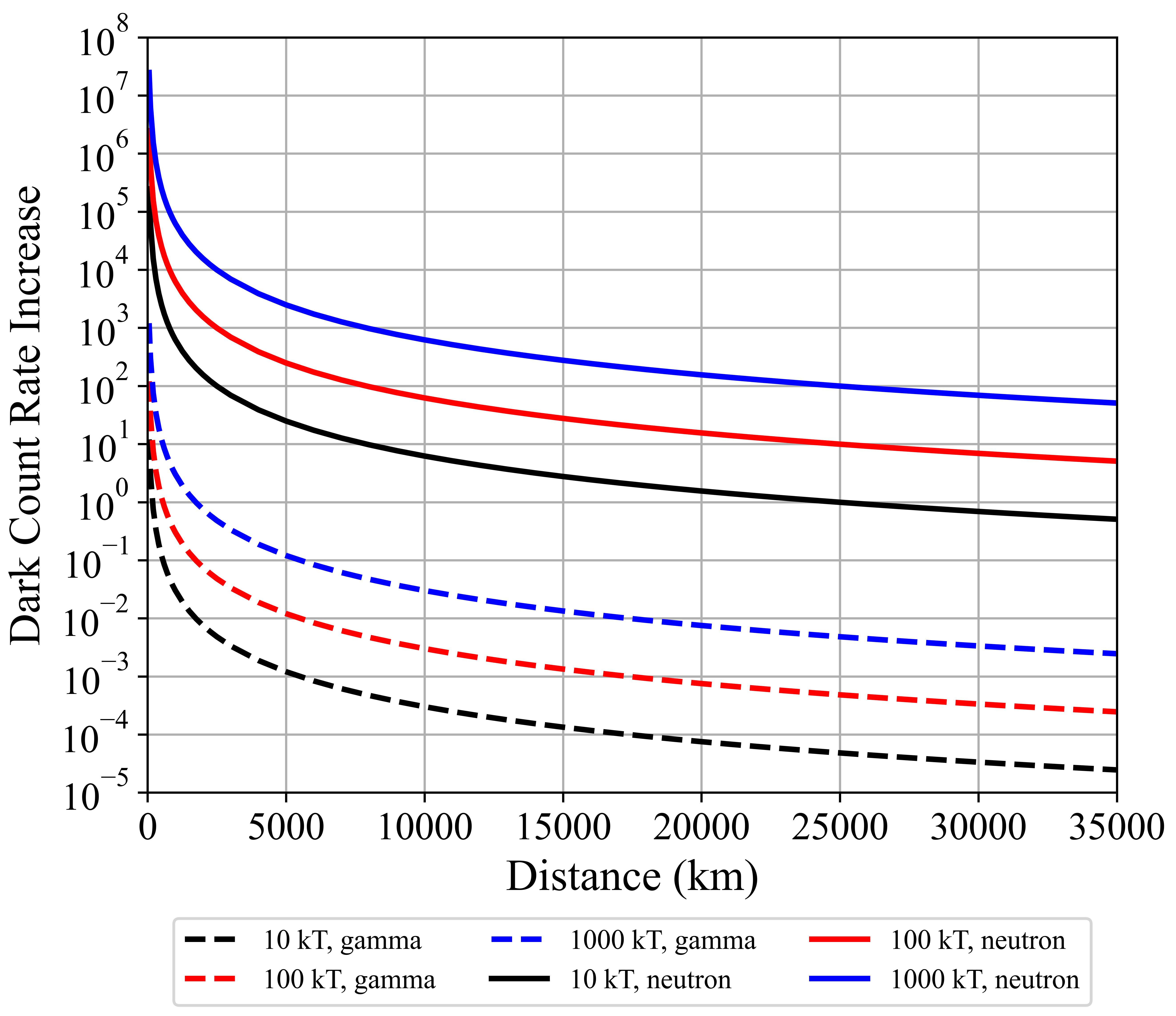}
	\captionsetup{justification=raggedright,
					singlelinecheck=false }
	\caption{DCR increase in a silicon SPAD from the prompt radiation emitted from a HANE event.}
	\label{fig:12}
\end{figure}
The neutron spectrum used in this calculation is a fast Watt spectrum for \textsuperscript{235}U taken from the nuclear code package SCALE \cite{scale2020} and the gamma ray spectrum is a \textsuperscript{235}U prompt gamma fission source also taken from SCALE \cite{scale2020}. Upon determining the NIEL damage imposed by the neutrons and gammas, the DCR increase is estimated using the conversion factor 36.26 $(MeV~cm^{2}~g^{-1})$ per 1 dark count per second increase at an operating temperature of -10~$C^{\circ}$ for a silicon SPAD. Fig. \ref{fig:12} shows the estimated DCR increase in the detector. Shielding is neglected for this calculation as 10~mm of aluminum shielding has a negligible effect on either flux. The DCR from neutrons is greater than the DCR from gamma rays due to the relatively large amount of NIEL damage that neutrons inflict on the silicon detector compared to gamma rays. Assuming that a HANE event occurs at an altitude of 100 km, the MEO and GEO satellites' detectors experience relatively low damage from the neutrons and gammas due to the $\frac{1}{r^2}$ dependence of the prompt radiation.

\begin{figure}[ht]
	\includegraphics[width=.45\textwidth]{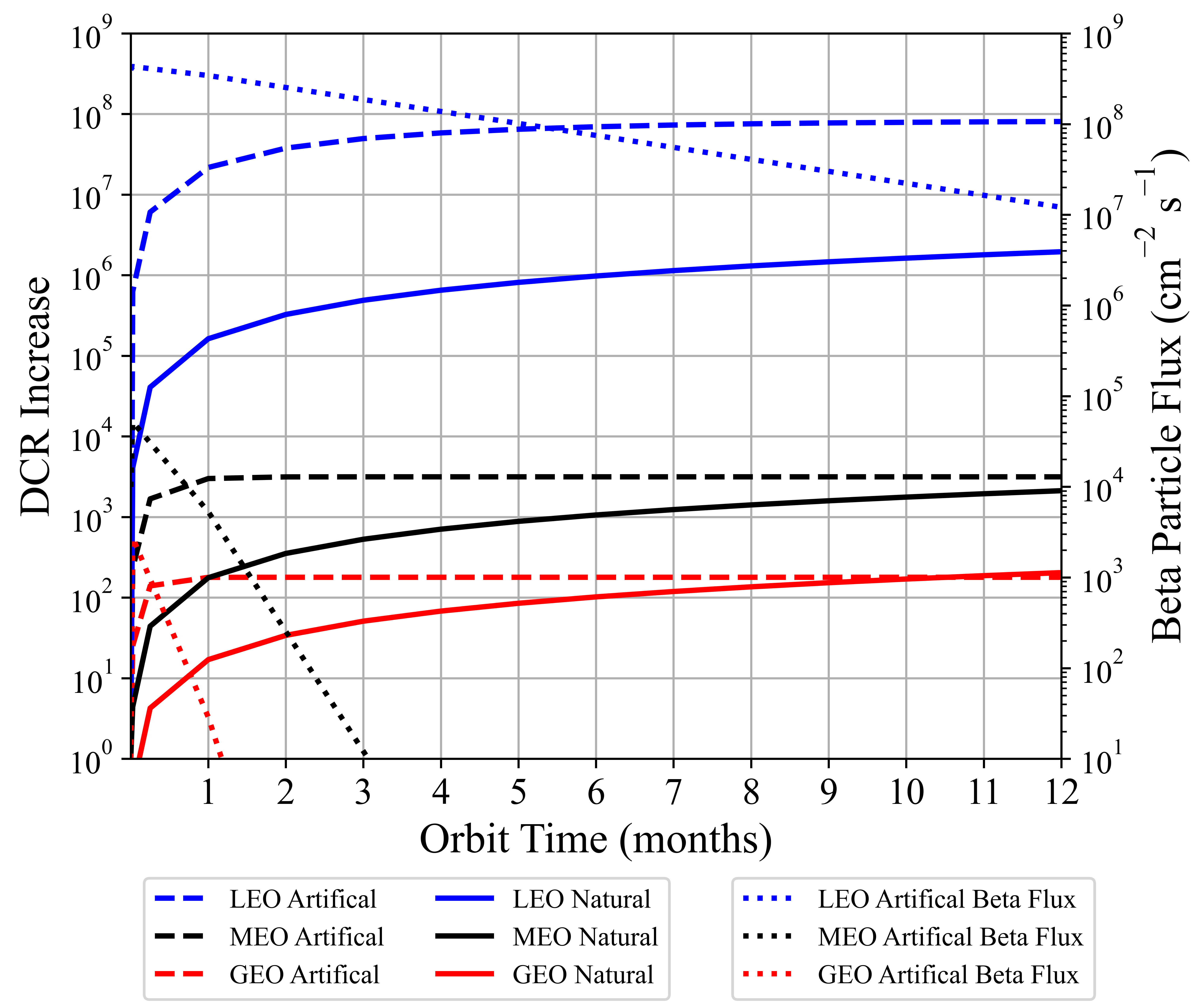}
	\captionsetup{justification=raggedright,
					singlelinecheck=false }
	\caption{DCR increase in a silicon SPAD from an artificial radiation belt created by a HANE for different satellite orbits.}
	\label{fig:13}
\end{figure}

The creation of an artificial radiation belt from a HANE can cause the radiation dose to a satellite to increase significantly. The high energy beta particles in the artificial belt penetrate more than the natural Van Allen Belt beta particles and produce higher energy bremsstrahlung radiation. The beta particle flux for the artificial radiation belts and their spectrum calculated previously (Table \ref{tab:rad_belt} and Fig. \ref{fig:5}) are combined with the mean lifetime of the electrons in their respective belts, to calculate the beta particle fluence that a satellite orbiting in an artificial radiation belt would receive. Upon calculating the fluence, we apply the shielding calculations from Fig. \ref{fig:10} to determine the NIEL damage that the beta particles and bremsstrahlung cause in the silicon SPAD detector. Converting the NIEL damage to a DCR increase, the DCR increase from the artificial belts on the different satellites, can be seen in Fig. \ref{fig:13}.

The combination of the higher beta particle fluxes and longer lifetimes of the electrons in the artificial belts at smaller magnetic latitudes creates a severe hazard for LEO satellite-based SPADs. The DCR increases significantly within the first 1 to 2 days and continues to climb over the next 6 months. The lifetime of the electrons in the MEO and GEO belts are on the order of 10 and 5 days, respectively, and after 1 month, the flux in these belts has already dropped by 1 to 2 orders of magnitude. This is seen prominently in Fig. \ref{fig:13} where the DCR increase in the detectors levels off after the first month. In the natural Van Allen Belts, the beta particles and bremsstrahlung swap as the leading contributor to the DCR increase when the shielding is increased to 10 mm of aluminum, but that is not the case for the artificial belts. The penetrating power of the high energy beta particles puts the beta particle and bremsstrahlung damage 3 orders of magnitude apart. Both the prompt radiation and the artificial radiation belts from a HANE adversely affect detectors on a LEO satellite, while the MEO and GEO satellites accumulate only minimal damage from both effects in comparison.

\begin{table*}
    \centering
     \caption{Increase in dark count rates in a single-photon avalanche detector onboard a satellite for various operating temperatures after 1 year, assuming a 1 MT detonation at 100 km.}
 \begin{tabular}{|c||c|c||c|c||c|c|}
 \hline
Satellite (Al Shielding) & LEO (1 mm) & LEO (10 mm) & MEO (1 mm) & MEO (10 mm) & GEO (1 mm) & GEO (10 mm) \\
\hline
\hline
\multicolumn{7}{|c|}{Natural Radiation Environment Only} \\
\hline
-10~$C^{\circ}$ & $7.39\times 10^{6}$ & $1.96\times 10^{6}$ & $7.49\times 10^{7}$ & $2.13\times 10^{3}$ & $6.10\times 10^{6}$ & $2.05\times 10^{2}$ \\
-40~$C^{\circ}$ & $3.96\times 10^{5}$ & $1.05\times 10^{5}$ & $4.01\times 10^{6}$ & $1.14\times 10^{2}$ & $3.27\times 10^{5}$ & $1.10\times 10^{1}$ \\
-60~$C^{\circ}$ & $5.62\times 10^{4}$ & $1.49\times 10^{4}$ & $5.69\times 10^{5}$ & $1.62\times 10^{1}$ & $4.64\times 10^{4}$ & $1.56\times 10^{0}$ \\
-80~$C^{\circ}$ & $7.98\times 10^{3}$ & $2.12\times 10^{3}$ & $8.09\times 10^{4}$ & $2.30\times 10^{0}$ & $6.60\times 10^{3}$ & $2.21\times 10^{-1}$
\\
\hline
\multicolumn{7}{|c|}{Natural Radiation and Prompt Nuclear Radiation} \\
\hline
-10~$C^{\circ}$ & $7.64\times 10^{6}$ & $2.21\times 10^{6}$ & $7.49\times 10^{7}$ & $2.28\times 10^{3}$ & $6.10\times 10^{6}$ & $2.56\times 10^{2}$ \\
-40~$C^{\circ}$ & $4.09\times 10^{5}$ & $1.18\times 10^{5}$ & $4.01\times 10^{6}$ & $1.22\times 10^{2}$ & $3.27\times 10^{5}$ & $1.37\times 10^{1}$ \\
-60~$C^{\circ}$ & $5.81\times 10^{4}$ & $1.68\times 10^{4}$ & $5.69\times 10^{5}$ & $1.74\times 10^{1}$ & $4.64\times 10^{4}$ & $1.95\times 10^{0}$ \\
-80~$C^{\circ}$ & $8.26\times 10^{3}$ & $2.39\times 10^{3}$ & $8.09\times 10^{4}$ & $2.47\times 10^{0}$ & $6.60\times 10^{3}$ & $2.77\times 10^{-1}$
\\
\hline
\multicolumn{7}{|c|}{Natural Radiation and HANE Artificial Belts} \\
\hline
-10~$C^{\circ}$ & $5.80\times 10^{9}$ & $8.29\times 10^{7}$ & $7.50\times 10^{7}$ & $4.11\times 10^{3}$ & $6.11\times 10^{6}$ & $3.17\times 10^{2}$ \\
-40~$C^{\circ}$ & $3.11\times 10^{8}$ & $4.44\times 10^{6}$ & $4.02\times 10^{6}$ & $2.20\times 10^{2}$ & $3.27\times 10^{5}$ & $1.70\times 10^{1}$ \\
-60~$C^{\circ}$ & $4.41\times 10^{7}$ & $6.30\times 10^{5}$ & $5.71\times 10^{5}$ & $3.13\times 10^{1}$ & $4.65\times 10^{4}$ & $2.41\times 10^{0}$ \\
-80~$C^{\circ}$ & $6.27\times 10^{6}$ & $8.96\times 10^{4}$ & $8.11\times 10^{4}$ & $4.44\times 10^{0}$ & $6.60\times 10^{3}$ & $3.43\times 10^{-1}$ \\
\hline
\multicolumn{7}{|c|}{Natural Radiation, Prompt Nuclear Radiation and HANE Artificial Belts} \\
\hline
-10~$C^{\circ}$ & $5.80\times 10^{9}$ & $8.31\times 10^{7}$ & $7.50\times 10^{7}$ & $4.26\times 10^{3}$ & $6.11\times 10^{6}$ & $3.68\times 10^{2}$ \\
-40~$C^{\circ}$ & $3.11\times 10^{8}$ & $4.45\times 10^{6}$ & $4.02\times 10^{6}$ & $2.28\times 10^{2}$ & $3.27\times 10^{5}$ & $1.97\times 10^{1}$ \\
-60~$C^{\circ}$ & $4.41\times 10^{7}$ & $6.32\times 10^{5}$ & $5.71\times 10^{5}$ & $3.24\times 10^{1}$ & $4.65\times 10^{4}$ & $2.80\times 10^{0}$ \\
-80~$C^{\circ}$ & $6.27\times 10^{6}$ & $8.98\times 10^{4}$ & $8.11\times 10^{4}$ & $4.61\times 10^{0}$ & $6.60\times 10^{3}$ & $3.98\times 10^{-1}$ \\
\hline
\end{tabular}
 \label{tab:f5_ray_tracing}
\end{table*}

\section{Discussion}
\label{sec:5}
The complexity of conducting this study leads to uncertainty in the final answers as there are numerous stochastic variables where simplifying assumptions were made. The natural radiation space environment is time and solar activity dependent and there is uncertainty in the models we use (AE8, AP8) that average the radiation spectrum and fluence a satellite would incur during orbit. The NIEL values are well documented in silicon devices but their correlation to dark counts in silicon SPADs are less certain. Each detector may vary and the impact of gamma rays and other particles on the DCR has not been experimentally well established in the literature. The artificial radiation belts from a HANE event are the least understood and so the results presented in this work are a first approximation and more research is needed to improve the accuracy of the impact of the HANE models. Nonetheless, the results for the different radiation environments, satellite orbits, shielding and operating temperatures are compiled into Table \ref{tab:f5_ray_tracing}.

It is desirable to maximize the  fidelity of quantum state transmissions over a quantum communications link and so from Fig. \ref{fig:1}, the dark count rates for the detector need to be as low as possible (if we assume 30~dB optical link loss, the DCR has to be lower than $1\times 10^{5}$). For a LEO satellite in polar orbit at 800~km, cooling the detector down to below -60~$C^{\circ}$ with at least 10~mm of aluminum shielding is the optimal combination to reduce the DCR to maximize the fidelity. The protons that the LEO satellite encounter when passing through the South Atlantic Anomaly and Polar Cusps are the main contributor to the increase in the DCR. The design of the shielding needs to be optimized around placing the most material between the external space radiation and the detector. Feeding optical fibers from the receiver to a detector located near the center of the satellite could provide additional shielding to the detector and lower the high energy proton flux further. Different LEO orbits (we chose a polar orbit at 800~km for this study) at closer ranges or in an equatorial orbit can also reduce the amount of proton radiation that the satellite encounters. The Chinese quantum satellite \emph{Micius} is positioned in a polar orbit with a perigee of 488~km and apogee at 584~km \cite{liao2017satellite}. This orbit reduces the amount of proton radiation that the satellite encounters by roughly an order of magnitude compared to an 800~km polar orbit. \par
When prompt radiation from a HANE occurs near a LEO satellite, the DCR increases by $10^5$ if the detector is operating at -10~$C^{\circ}$. In all likelihood, if a LEO satellite receives prompt radiation from a HANE, the satellite is likely to be incapacitated from the x-ray, thermal and other effects of the nuclear detonation. In the event the HANE occurs at the correct geomagnetic latitude to create an artificial radiation belt in the orbit plane of a LEO satellite, the detector's DCRs will increase significantly causing a loss in the ability to perform quantum communications regardless of cooling or shielding. \par
Silicon SPADs on MEO and GEO satellites can incur less radiation damage than their LEO counterparts if the detectors are adequately shielded ($>$10~mm aluminum). The low energy protons in the outer Van Allen Belt can be easily shielded with one mm of aluminum while the beta particles can be almost completely shielded by 10~mm of aluminum. With 10~mm of aluminum shielding, the beta particles and protons have a minimal effect on the DCR, and the biggest contribution is from the bremsstrahlung radiation from the electrons stopping in the shield. As the  extra shielding does not stop high energy photons, further shielding has little benefit to detector performance. Our models have the beta particles encountering the aluminum shielding but, if a low atomic number material, like polyamide, is placed on the outside of the satellite, the bremsstrahlung radiation can be reduced further. In the event of a HANE, both MEO and GEO satellite detectors encounter only slight DCR increases from both the prompt radiation and the artificial radiation belts. The large distance of the MEO and GEO satellites from Earth drops the prompt radiation levels to a minimal amount and the large volume of a radiation belt reduces the beta particle flux to a manageable level. In regards to radiation induced dark counts in a SPAD, the MEO and GEO satellite orbits provide the best platforms.

\section{Conclusions}
\label{sec:6}

Cooling and shielding a silicon single photon detector are the best ways to mitigate the radiation hazards of space, but both come with trade-offs in terms of cost, size, and complexity. For detectors on LEO satellites, protons are the biggest issue in terms of radiation damage and are best mitigated by choosing an orbit that minimizes exposure when passing through the South Atlantic Anomaly and Polar Cusps, as well as by placing the detector in the middle of the satellite. SPADs on MEO and GEO satellites, if shielded by more than 10 mm of aluminum, provide the best platform when measured in terms of the least amount of radiation damage to the detectors. In the event of a HANE, the artificial radiation belts produced by the explosion  will likely cause too much damage to silicon SPADs in LEO and render them unfit for quantum communications.  At the same time, MEO and GEO satellite orbits should only suffer minor DCR increases.

\begin{acknowledgments}
This work was supported by Defense Threat Reduction Agency Award HDTRA1-93-1-201.  We thank Brian P. Williams for discussions.
\end{acknowledgments}
\bibliography{dcr.bib}

\end{document}